# Quantum Imaging for Semiconductor Industry

**Anna V. Paterova**[1,*]**, Hongzhi Yang**[1]**, Zi S. D. Toa**[1]**, Leonid A. Krivitsky**[1, **]

[1]*Institute of Materials Research and Engineering,*

*Agency for Science Technology and Research (A*STAR), 138634 Singapore*

*[Paterova_Anna@imre.a-star.edu.sg](mailto:Paterova_Anna@imre.a-star.edu.sg)

**[Leonid_Krivitskiy@imre.a-star.edu.sg](mailto:Leonid_Krivitskiy@imre.a-star.edu.sg)

## Abstract

Infrared (IR) imaging is one of the significant tools for the quality control measurements of fabricated samples. Standard IR imaging techniques use direct measurements, where light sources and detectors operate at IR range. Due to the limited choices of IR light sources or detectors, challenges in reaching specific IR wavelengths may arise. In our work, we perform indirect IR microscopy based on the quantum imaging technique. This method allows us to probe the sample with IR light, while the detection is shifted into the visible or near-IR range. Thus, we demonstrate IR quantum imaging of the silicon chips at different magnifications, wherein a sample is probed at 1550 nm wavelength, but the detection is performed at 810 nm. We also analyze the possible measurement conditions of the technique and estimate the time needed to perform quality control checks of samples.

Infrared (IR) metrology plays a significant role in areas ranging from biological imaging[1-5] to materials characterisation[6] and microelectronics[7]. For example, the semiconductor industry utilizes IR radiation for non-destructive microscopy of silicon-based devices. However, IR optical components, such as synchrotrons[8], quantum cascade lasers, and HgCdTe (also known as MCT) photodetectors, are expensive. Some of these components, in order to maximize the signal-to-noise ratio, may require cryogenic cooling, which further increases costs over the operational lifetime. Furthermore, IR devices are often subjected to export control regulations, adding extra burden for industrial system integrators. Thus, it is attractive to seek alternative approaches of implementing IR metrology without such expensive optical components.

One of the approaches is based on nonlinear interferometry[9-11] using pairs of spectrally nondegenerate entangled photons, produced via spontaneous parametric down-conversion (SPDC). In this process, the laser photon in a nonlinear crystal decays into a pair of correlated photons called signal and idler, which have visible and IR wavelengths, respectively. The photons are spatially split, so that the IR photon interacts with the sample, while the visible photon is used as a reference. All the photons propagate through the nonlinear crystal a second time, during which another SPDC process takes place. Due to quantum interference between the photon pairs, the interference pattern of the visible photons, generated during these two passes of the laser beams through the crystal, reports on the properties of idler photons that interacted with the sample. In this way, IR metrology can be carried out using inexpensive visible optical components, such as inexpensive CCD or CMOS cameras.

Earlier, the concept of nonlinear interferometry was applied to a number of metrological applications, including imaging[12-14], spectroscopy[15-21], optical coherence tomography[22,23], and polarimetry[24]. In this work, we demonstrate its application to the imaging of actual microelectronic devices from our industrial partner for quality inspection and control of the fabrication process.

The schematic of our nonlinear interferometer is shown in Fig. 1. Detailed discussions can be found in earlier works[13,19,23]. Thus, we limit ourselves to a brief description here. A portion of the pump laser beam is down-converted in a nonlinear crystal into a pair of frequency non-degenerate entangled photons. The phase matching condition in the crystal is set in such a way that one photon of a pair is generated in the visible and another in the IR range. The IR photons are subsequently split from the pump and visible photons using a dichroic beam-splitter. Visible and pump photons are reflected back into the crystal by a mirror, while IR photons are reflected by the sample under study. The reflected pump beam generates another pair of photons, identical to those generated in the first pass through the crystal. The wave-functions of the two photon pairs, coherently generated in the two passes of the pump through the crystal, interfere. The modulation of the intensity of signal photons is defined by the phases of all the interacting waves, while the contrast (visibility) of interference fringes depends on the reflectivity of the sample placed in the IR arm of the interferometer. Assuming the spatial structure of the sample in the IR arm, one can write[13,23]:

$$I_s(x,y) \propto 1 + |\mu(x,y)||\tau_i(x,y)|^2|r_i(x,y)|\cos(\varphi_p - \varphi_s(x,y) - \varphi_i(x,y)), \quad (1)$$

where $\tau_i$ and $r_i$ are the amplitude transmission and reflection coefficients of the idler photons through the nonlinear interferometer, respectively, $\mu$ is the normalized first order correlation function of SPDC photons, and $\varphi_{p,s,i}$ are the phases of the pump, signal and idler photons, respectively.

Let us now continuously vary the relative phase $\varphi_i = [0, \pm 2\pi n]$ between the interferometer arms (by scanning the sample position) and calculate the deviation of the signal from its mean value (or also called standard deviation (STD), $\sigma(x,y)$):

$$\sigma(x,y) \sim V(x,y)/\sqrt{2}, \quad (2)$$

where $V(x,y) = |\tau_i(x,y)|^2|r_i(x,y)|$ is the visibility of the interference fringes for the balanced interferometer $|\mu| = 1$. Thus, the variance, defined as the square of the STD value, of the interference fringes is equivalent to the reflectivity of the sample:

$$Var(x,y) = \sigma^2(x,y) \sim |r_i(x,y)|^2 = R_i. \quad (3)$$

The experimental setup of the method is shown in Fig. 1. We use a continuous wave single-mode 532 nm laser (Coherent, ~16 mW power after fiber) as a pump for the periodically poled lithium niobate (PPLN) crystal (HC Photonics Corp.), in which SPDC occurs. The radius of the pump beam is reduced and collimated before the dichroic beamsplitter $D_1$ (Semrock), which reflects the pump towards the PPLN crystal. Note that the pump spatial structure is preserved during the two-way travel through the interferometer. The PPLN crystal has a 7.5 μm poling period, and is mounted on a temperature controlled stage kept at a temperature of 70 °C with stability better than 0.1 °C. Under these conditions, visible and IR photon pairs are generated at 810 nm and 1550 nm wavelengths, respectively. Using a dichroic beam splitter $D_2$ (Semrock), the pump and the visible beams are reflected while the IR beam is transmitted.

A system of three confocal lenses is placed in each arm of the interferometer. The focal lengths of the lenses correspond to $F_1$ = 75 mm, $F_2$ = 75 mm, and $F_3$ = 25 or 5 mm (Edmund Optics). The imaging system in each arm projects the de-magnified image of the angular spectra of the SPDC on the mirror and the sample, respectively.

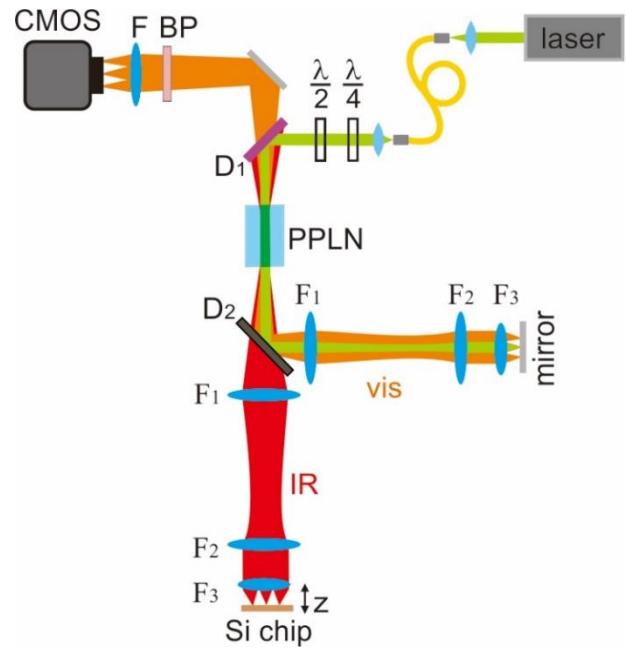

FIG. 1. Schematic of the experimental setup. A continuous wave laser generates non-degenerate SPDC photon pairs at the PPLN crystal. Photons are split into two arms of the Michelson interferometer according their frequencies: pump and signal photons reflected to the visible arm, and idler photons are transmitted into IR arm. Then, visible and IR photons are reflected back by mirror and Si chip, respectively. Due to the three-lens system the beams preserve their spatial spectrum. Next, the reflected pump generates SPDC photons a second time. The interference pattern of the visible SPDC photons is captured by focusing the signal with the lens F into the standard CMOS camera.

The visible and pump beams are reflected by a metallic mirror, while the IR beam is reflected by the sample. The sample is mounted on a kinematic mount and a piezo stage (Piezosystem Jena, Model: MIPOS 100 PL), which is used to set the imaging plane and scan the relative phase between interferometer arms $\varphi_i$. The pitch and yaw of the mirror and the sample are aligned such that the reflected beams travel back to the crystal. Once the arms of the interferometer are balanced, the interference of the visible beam is observed using off-the-shelf CMOS camera (Thorlabs, Model: CS2100M-USB). The signal is spectrally filtered by a band pass filter BP (Semrock, 810 nm, bandwidth = 10 nm).

In the experiment, we scan the position of the sample z within a range of few microns. We capture the interference pattern at each position of the sample with the 300 ms acquisition time. The obtained images are subsequently processed and the variance of the signal for each pixel of the camera is calculated. This results in a spatial map of the variance, which corresponds to the reflectivity profile of the sample, see Eq. (3).

First, we measure the spatial resolution of our system by imaging the metal coated microscope test slide. The results are shown in Fig. S1 of the Supplementary Materials. The achieved spatial resolution of our method is 39 μm and 7.8 μm, with $F_3$ = 25 mm and $F_3$ = 5 mm lenses, respectively. The resolution of the technique can be improved further by using lens $F_3$ with smaller focal length or by using solid immersion lens (SIL)[25].

Next, we use silicon chips, which are provided by our industrial partner, as samples for our study, see Figs. 2(a) and 3(a). The first sample in Fig. 2(a) comprises a pattern of structured copper contacts, coated on a silicon substrate. The contacts on the substrate are disjoint in initial stages of fabrication. The substrate is subsequently capped by another layer of silicon, which also has patterned copper contacts, see Fig. 3(a). The contacts on the top silicon layer are aligned with the contacts at the bottom substrate to form a continuous electrical circuit in the device. Any misalignment between the two silicon layers during the capping process leads to disconnected metal contacts and results in interruption of the current flow. This causes device failure. Thus, inspection of the contact connections after the capping is critical for improvements to the fabrication process. Since the top layer of the device is made of silicon, see Fig. 3(a), it is not possible to use visible light for the inspection. Instead, IR light has to be used.

White-light microscopy of the uncapped sample shows exposed copper contacts on the sample substrate, see Fig. 2(a). The image of the sample using our technique is shown in Fig. 2b. The image corresponds to the variance of the interference patterns, where the interferometer arm with the sample is moved from z=0 to 1550 nm with 20 nm step. (This shift corresponds to $\varphi_i$ = [0, 4π], see Fig. S2(d) in Supplementary Materials). The arrangement of the contacts imaged by our technique is found to coincide with that observed using white light microscopy.

Interestingly, the surface of the metal contacts appears dimmer than the surface of the surrounding silicon substrate. Upon closer inspection, we found that they have a granular appearance, which is due to strong scattering from the metal lines.

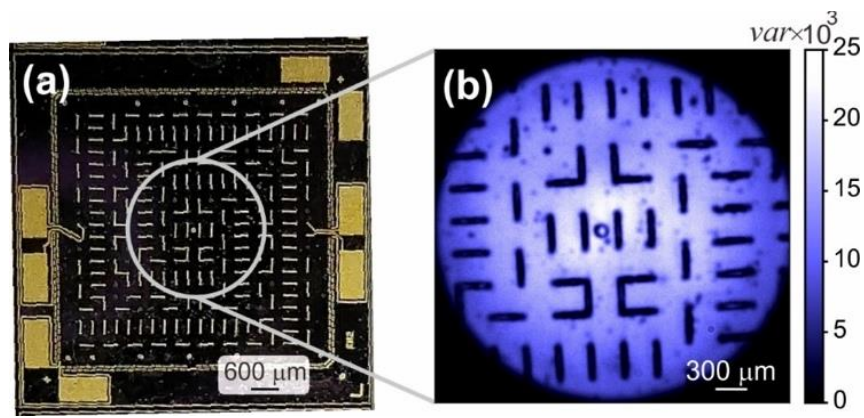

**FIG. 2.** (a) White-light microscope image of uncapped sample. (b) Quantum microscope image of uncapped sample with $F_3$ = 25 mm. The dark stripes are the metal contacts, which mostly scatter the probing IR light.

In principle, even one interference pattern at a given scanning position may be enough to see the structure of the sample. However, this is possible only if the sample is flat and the relative phase $\varphi_i$ is equal to $\pm 2\pi n$ or $\pm(2n+1)\pi$, see Fig. S2(a), (c). If the relative phase $\varphi_i$ is equal to $\pm(2n+1)\pi/2$, the full image of the sample is not observable, see Fig. S2(b). Therefore, we perform variance measurements of the interference pattern, where we do not need to specifically look for the exact relative phases $\varphi_i$.

Similarly, we perform imaging of the chip with the silicon layer on top, see Fig. 3(a). As expected, one is not able to see through the silicon layer using visible light. The quantum imaging results obtained at various magnifications, as we image through the silicon top layer, are shown in Figs. 3(b) and 3(c). As observed, our method allows imaging electrical interconnects through the silicon layer. Since the capping layer has the inverse pattern of the electrode, upon capping the electrodes connect and form a continuous path for electrical current. A magnified view of the connection area allows us to gauge the accuracy of the capping process, see Fig. 3(c). Interestingly, the through-silicon images show additional magnification, which is due to the immersion of the structure in the high refractive index of ~3.5 (at 1550 nm) for a silicon.

Note that there are dimmer parts of the image at the sides, which is probably due to bending of the silicon layer on top. This may result in appearance of artefacts in the observed images. However, imaging of the sample at higher magnification allows a closer inspection of such artefacts.

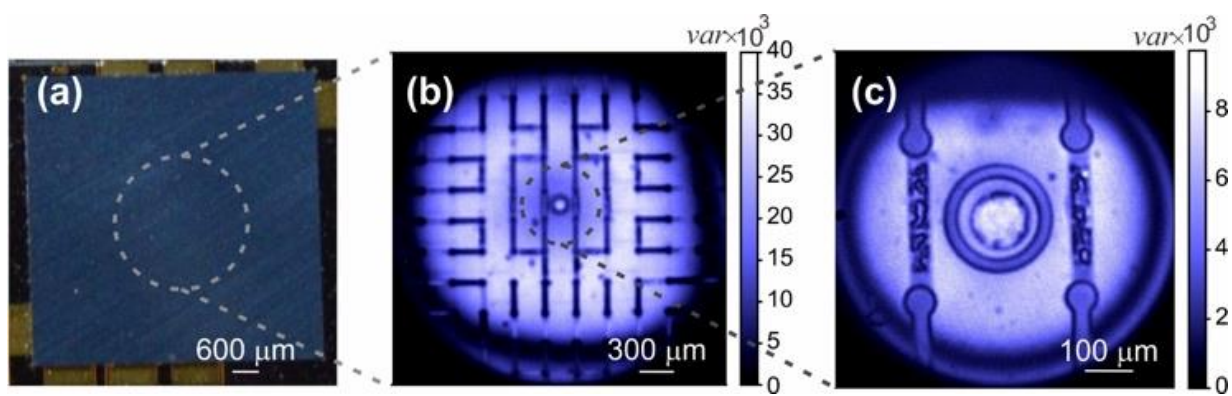

**FIG. 3.** (a) White-light microscope image of the capped sample. Quantum microscope image of the capped sample with (b) $F_3$ = 25mm and (c) $F_3$ = 5 mm lenses.

As it was mentioned earlier, in order to obtain the image of the samples, we perform the variance calculation for the interference patterns measured by continuous scanning of the relative phase $\varphi_i$. However, continuous scanning is an idealistic approach, as it may not be reproducible in the real factory conditions. Therefore, the possibility of employing compressive sensing[26] in our technique is also explored.

Additional perturbations may be introduced in the scanning of the sample during imaging. Therefore, to account for such perturbations (e.g. measurement of the interference pattern at the same position twice, or jumps in the scanning position more than $2\pi$), we take the data corresponding to $\varphi_i = [0, 4\pi]$ and select a number of measured interference patterns to calculate the variance of the signal. Figure 4(a) shows the variance for the half of the data with $\varphi_i = [0, 2\pi]$. The result is similar to the image in Fig. 3(c), but noisier. Figures 4(b)-(d) correspond to variance of randomly selected 10, 5 and 2 interference patters, respectively. Details in the imaged field of interest are still discernable despite discarding more than half of interference patterns recorded in the measurement. In particular, fine details in the alignment of the metal contacts are discernable with 10 interference patterns, see Fig. 4(b). This suggests that random sampling at 10 positions with cheap piezo controllers is adequate for noisy but detailed imaging of the silicon chips. The image is also qualitatively accurate with the reduction to 5 interference patterns, see Fig. 4(c) and Fig. S3 in Supplementary Materials. However, the image is barely discernable with the variance of only 2 interference patterns, see Fig. 4(d).

These results show that capturing of 5 random interference patterns with measurement time t=1.5 s is enough to acquire an image of a given sample (see also Fig. S4 of the Supplementary Materials). However, the measurement time can be further decreased by using higher power laser and longer nonlinear crystals, or by implementing high gain[27] or stimulated[28] regimes of SPDC generation.

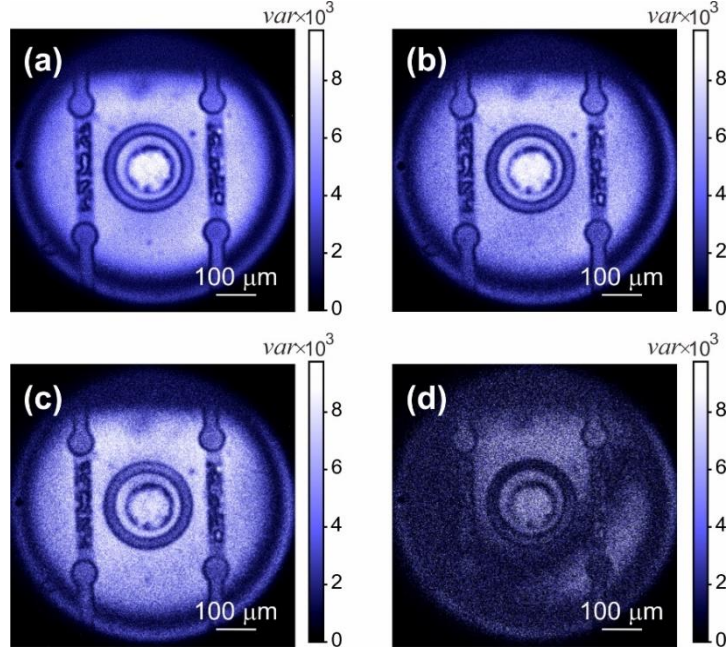

**FIG. 4.** Quantum microscope images taken with $F_3$ = 5 mm lens. The images correspond to variance of (a) half of the measured data, and randomly selected (b) 10, (c) 5, and (d) 2 interference patterns in increasing order of noise (or deterioration).

Thus, we believe that a compressive sensing[26] can be deployed in our technique. This allows for on-the-fly characterization of microelectronics chips and subsequent real-time feedback into the fabrication process.

In conclusion, we demonstrated wide-field NIR imaging, where the probing wavelength is at 1550 nm, while detection occurs at 810 nm. We show the application of the technique to the observation of real microelectronics structure concealed by a layer of opaque material in the visible range. The technique provides adequate spatial resolution, image quality and readout speed. Its main benefit is the use of simple and accessible components with visible light. In addition, the developed technique is intrinsically interferometric. Thus, it may reveal additional features, which may not be visible in a standard microscopy, such as material stress, defects and non-uniformity. We foresee that it has potential for broad adoption in the microelectronics industry for quality assurance and control.

## Supplementary Material

See supplementary material for optical resolution of the quantum imaging setup, phase data, variance calculation of four sets of five randomly selected interference patterns, and quantitative analysis of the compressive sensing approaches.

## Acknowledgements

We acknowledge the support of A*STAR "Quantum technology for the engineering" (QTE) program. We are grateful to staff of Kulicke and Soffa Industries, Inc. for providing the sample and encouraging discussions.

## Data availability

The data that support the findings of this study are available from the corresponding author upon reasonable request.

# Supplementary Materials for:
# Quantum Imaging for Semiconductor Industry

**Anna V. Paterova[1,*], Hongzhi Yang[1], Zi S. D. Toa[1], Leonid A. Krivitsky[1]**

[1]*Institute of Materials Research and Engineering, Agency for Science Technology and Research (A\*STAR), 138634 Singapore*

*Paterova_Anna@imre.a-star.edu.sg

**Leonid_Krivitskiy@imre.a-star.edu.sg

## Resolution of the quantum imaging setup

We measure the resolution of the technique using a negative USAF1951 resolution test target (Thorlabs). The images of the resolution test target with $F_3$=25 mm and $F_3$=5 mm lenses are shown in Fig. S1(a) and S1(b), respectively. Following the Rayleigh criterion of the resolution, we find that our method can perform imaging with 39 μm (for $F_3$=25 mm lens) and 8 μm (for $F_3$=5 mm lens) resolution.

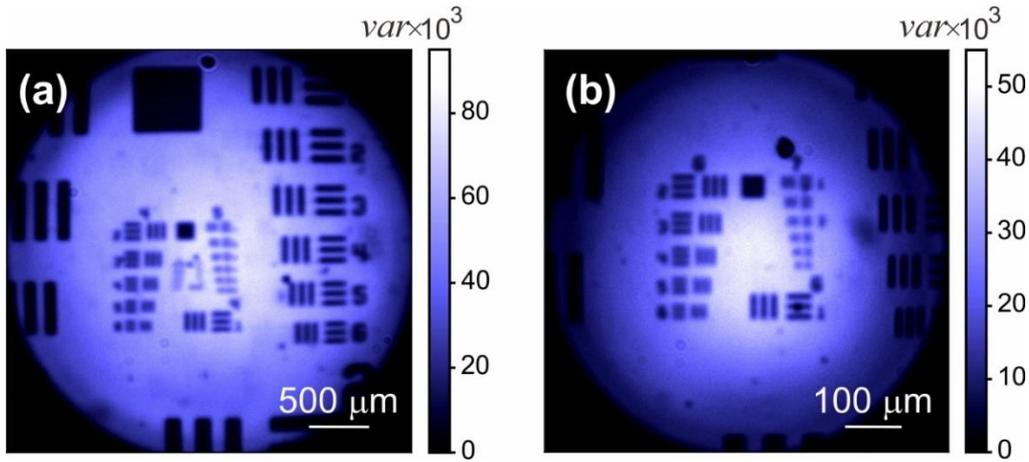

**FIG. S1.** Resolution measurement of the technique for (a) $F_3$ = 25 mm and (b) $F_3$ = 5 mm lenses.

## Phase data

In our experiments we capture the interference pattern with the CMOS camera at every step of the piezo positioner. Interference patterns at z=0.37 μm, 0.47 μm and 0.68 μm are shown in Figs. S2(a), S2(b) and S2(c), respectively. Figure S2(a) corresponds to the interference pattern, when the relative phase $\varphi_i=\pi$. According to Eq. (1), in this case the interference presents its minimum. Here, only the metal contacts appear brighter, as these regions do not present the interference due to scattering.

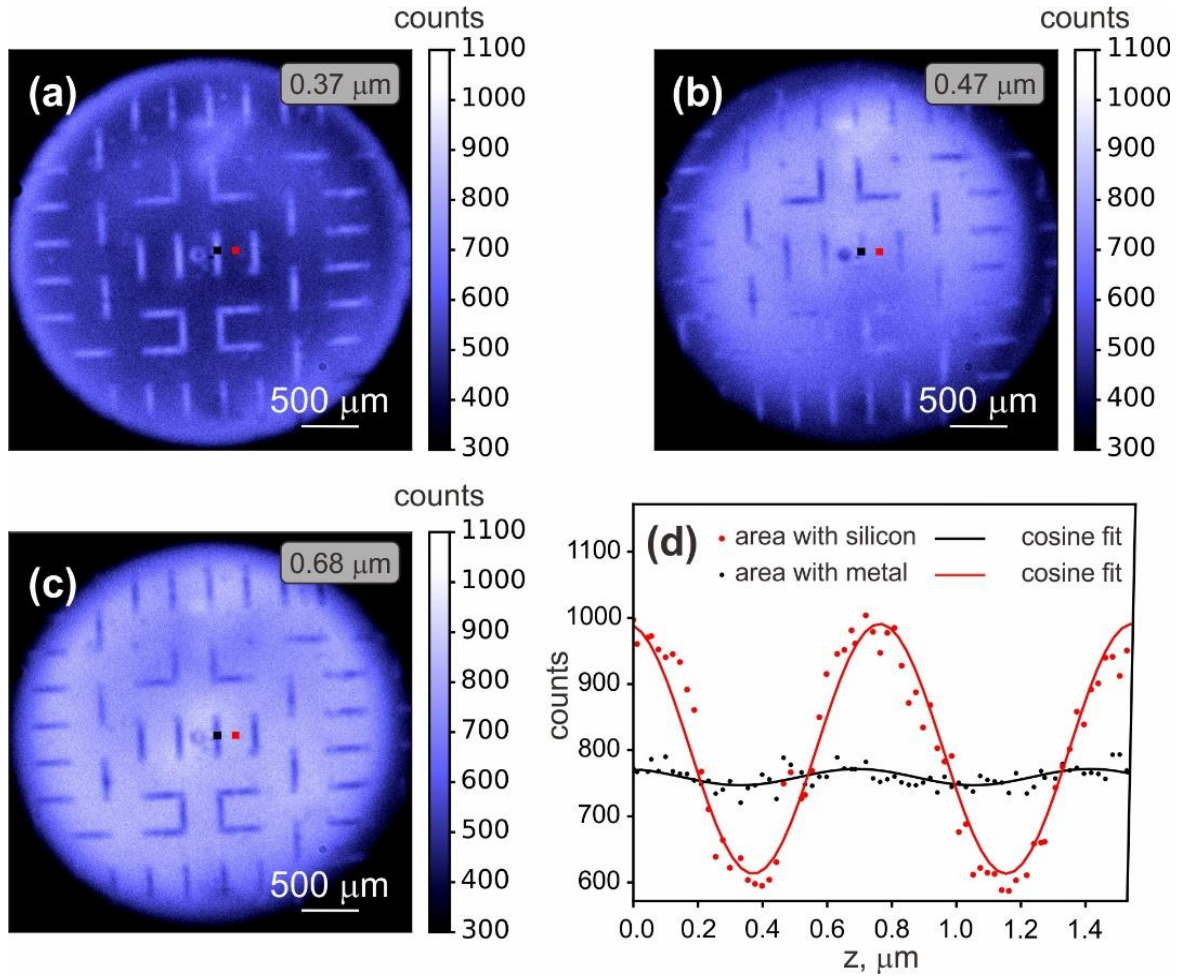

**FIG. S2.** Interference pattern captured by CMOS camera at (a) z=0.37 μm, (b) z=0.47 μm and (c) z=0.68 μm position of the piezo stage. (d) Interference fringes obtained for the different pixel areas at CMOS camera. Black and red points represent the data corresponding to metal contact and silicon area, respectively. Solid curves show the cosine fit of the data.

Figure S2(b) shows the resulting interference pattern for $\varphi_i=\pi/2$. In this case the interference pattern shows its mean value. However, due to the non-flat surface of the sample, we still can see the shape of some metal contacts. Figure S2(c) corresponds to $\varphi_i=0$, when the interface pattern shows its maximum. Here metal contacts appear darker, opposite to the case in Fig. S2(a). Figure S2(d) shows signal on the CMOS camera pixels depending on the shift of the piezo stage. The selected pixels are indicated by black and red points in Figs. S2(a), (b) and (c).

## Variance calculation for a random set of data

Figure S3 shows the variance of the five randomly selected interference patterns from the measured data (64 interference patterns). Each picture in Fig. S3 corresponds to different randomly selected sets of interference patterns to be included in the variance calculation. For each run, the calculated variance is slightly different. However, the qualitative image of the sample remains the same.

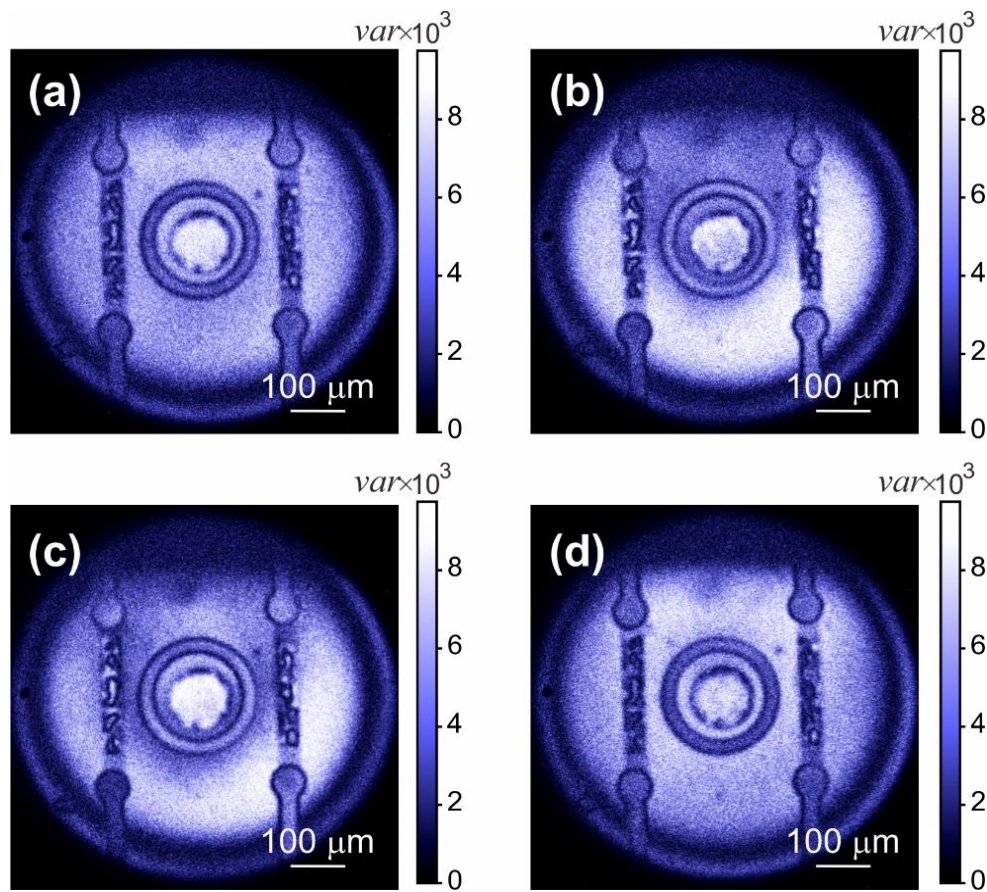

**FIG. S3.** (a)-(d) Four runs of the variance calculation for five randomly selected interference patterns from the measured data.

## Quantitative analysis of the compressive sensing approaches

Compressive sensing[S1] in our technique is explored (Fig. S4). This is done via the following three approaches. One, a continuous sequence of 32 or 64 interference patterns is selected from one phase oscillation in the full dataset. This is similar to scanning from z=0 to 770 nm or 1550 nm of the sample position in 20 nm step. Two, gaps in the sample position are introduced by discarding one to seven interference fringes between ones to be included in calculation. This is similar to shifting the interferometer arm from z=0 to 1550 nm in 40 to 140 nm steps. Three, a decreasing number of interference patterns are randomly selected from the full dataset.

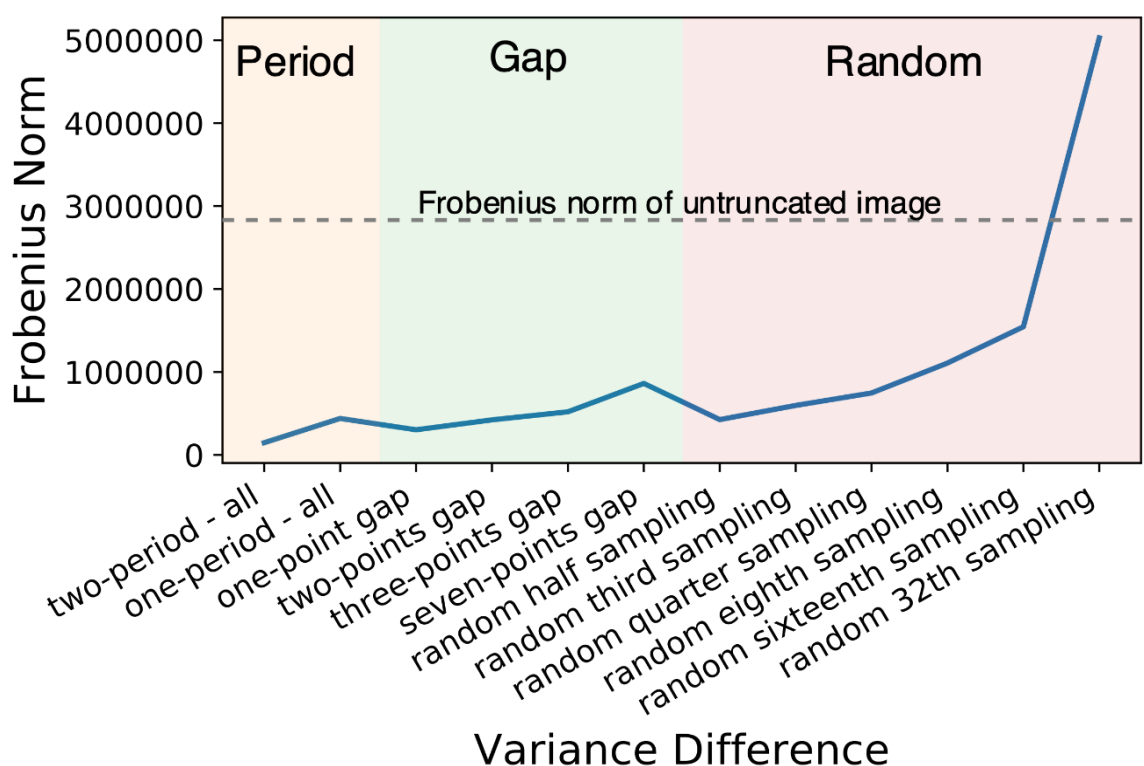

**FIG. S4.** Frobenius norms of the differences between the image calculated from variance of all interference fringes in a measurement, see Fig. 3(c), and different extents of truncation. These ranges from one and two phase periods in the visible beam, one to seven consecutive discarded fringes to randomly selected fringes as a decreasing fraction of the full dataset. For reference, the Frobenius norm of an untruncated image is ~2800000 and shown as the dotted line.

A single-valued metric of the difference between the images (or arrays) obtained from a variance calculation of the full dataset and those from truncated datasets is desirable for quantitative analysis. Matrix norms of the difference between two matrices are commonly used to quantify the "distance" between two matrices. One such matrix norm is the Frobenius norm, which is defined to be the square root of the sum of the squares of the individual elements in a matrix. It is chosen as the single-valued metric, as it is fast and easy to compute and widely used in optimization problems[S2-S4]. Mathematically, the Frobenius norm is expressed as:

$$\|A\|_F = \sqrt{\sum_i \sum_j |a_{ij}|^2}$$

where $a_{ij}$ is an element in the matrix $A$.

A value of zero for the Frobenius norm of the difference between two images would thus imply a lack of difference between elements in the two images. Furthermore, with this single-valued metric, the truncation of the dataset for the variance calculations can be considered as the introduction of noise into the data, as demonstrated by the increased pixelation or granularity in the variance images obtained from increasingly truncated datasets (Fig. 4).

The Frobenius norm (of the difference between untruncated and the various truncated datasets) generally increases with increasing truncation (Fig. S2). Note that the Frobenius norm of the untruncated dataset is ~2800000. Frobenius norms of the differences up to one-eighth random sampling (10 interference patterns) are less than half of that of the untruncated dataset. Note that continuous sequences of interference patterns (one and two phase periods) are generally less noisy than gapped or randomly selected datasets. It is thus very attractive from an industrial standpoint that an image (especially the alignment of the metal contacts) is already discernable with one-sixteenth of the sampling (5 interference patterns) conducted in one full measurement, see Fig. 4(c). Thus, this demonstrates that significant time can be saved in sample measurement, which allows for high throughput.